\documentstyle[12pt,epsf]{article}
\begin{document}

\begin{centering}

\bf \large 
PSEUDOGAPS AND CHARGE BAND
IN THE PARISI SOLUTION OF
INSULATING AND SUPERCONDUCTING ELECTRONIC 
SPIN GLASSES AT ARBITRARY FILLINGS

\end{centering}

\centerline{Reinhold Oppermann and Heiko Feldmann}

\begin{flushright}
\begin{minipage}{16.5cm}
\small {\bf Abstract.} We report progress in understanding the fermionic Ising
spin glass with arbitrary filling. 
A crossover from a magnetically disordered single band phase via two
intermediate bands just below the freezing temperature to a 3-band
structure at still lower temperatures -
beyond an almost random field instability - is shown to emerge in the
magnetic phase. 
An attempt is made to explain the exact solution in
terms of a quantum Parisi phase. 
A central nonmagnetic band is found and seen to become sharply separated
at $T=0$ by gaps from upper and lower magnetic bands. The gap sizes tend
towards zero as the number of replica symmetry breaking steps increases
towards infinity. In an extended model, the competition between local
pairing superconductivity and spin glass order is discussed.
\end{minipage}
\end{flushright}

{\bf 1. INTRODUCTION}\\
The fermionic Ising spin glass has many interesting and important
features. To mention a few:\\
i) It is an insulating limit of a large class of itinerant models that
involve frustrated magnetic order,\\
ii) its classical part coincides at $T=0$ with the $S=1$ 
Ghatak-Sherrington model [1], whose nonmagnetic
$S=0$-section differs slightly at finite temperatures,\\
iii) its definition on the Fock space allows quantum-dynamical
fermionic correlations in addition to the static spin- and
charge-correlations; \\
iv) the existence of low energy excitations in this sector of
quantum-dynamical correlations was seen to be connected with full
(infinite-step) replica permutation symmetry breaking of the Parisi
type [2]. The fascinating aspect was that infinite breaking
of discrete symmetries may lead to soft modes commonly known from weak
breaking of continuous symmetries;\\ 
v) its Onsager reaction field led to complications in the mean field
theory comparable to those encountered in the infinite-dimensional Hubbard
model [3],\\ 
vi) the strong coupling of charge and spin fluctuations causes a specific 
influence of frustrated magnetic interactions on
electronic transport including superconductivity [4].\\

\begin{figure}
\epsfxsize=8cm
\epsfbox{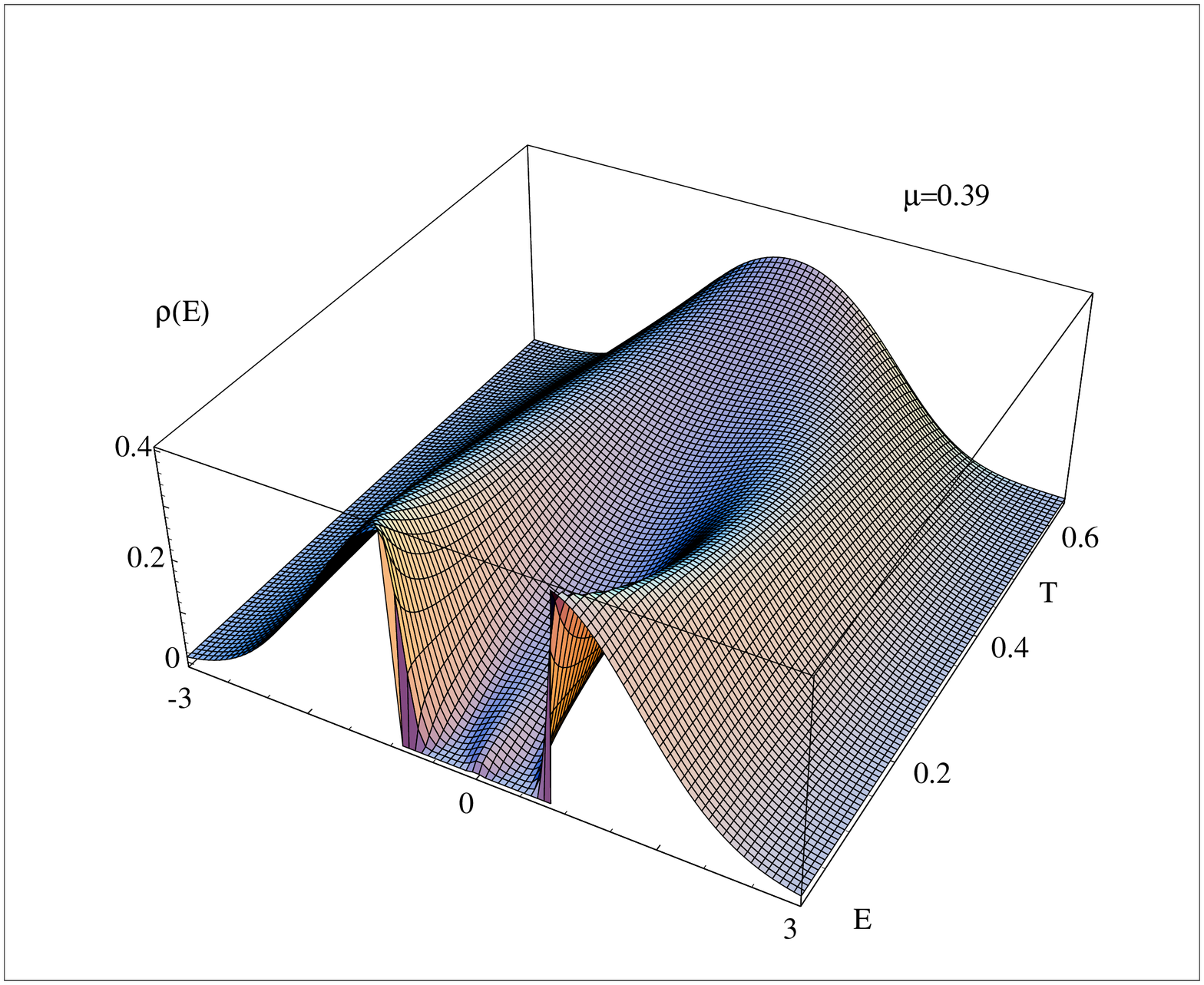}
\par\vspace{-6.6cm}\hspace{8cm}
\epsfxsize=8.3cm
\epsfbox{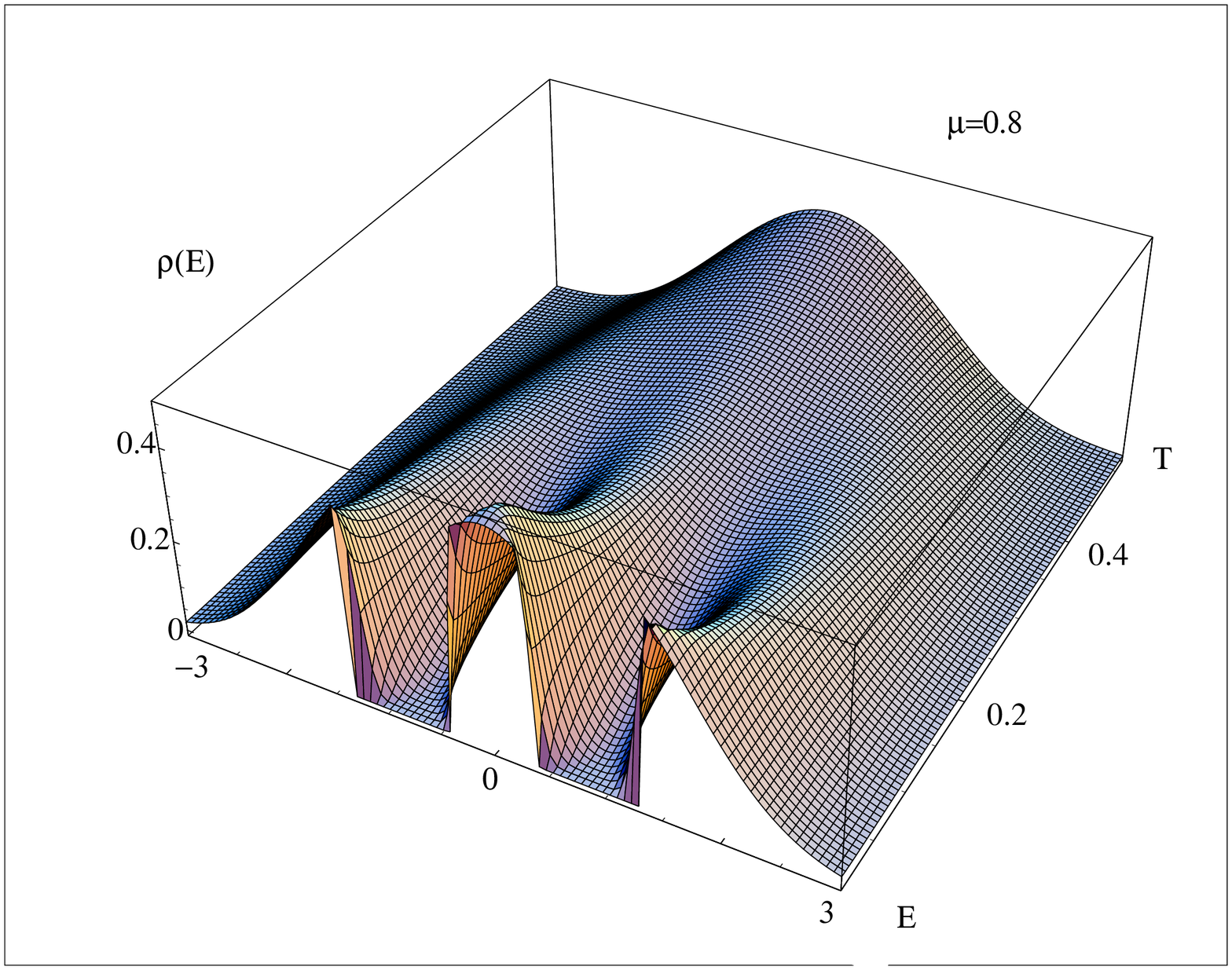}
\vspace{.5cm}
\caption{\small{Replica symmetric density of states below $T_f$ showing at
a small central peak emerge (and disappear at lowest $T$, left figure) for
a chemical potential $\mu=0.39 J$ just below the (0RPSB-) gap energy
$J\sqrt{2/\pi}$ and a large one for $\mu=0.8 J$ in the right figure}}
\label{dos}
\end{figure}

{\bf 2. MAGNETIC AND NONMAGNETIC BANDS IN THE INSULATING MODEL}\\
In order to obtain conclusions about the band structure of the model
we first solved analytically and numerically a set of coupled
selfconsistent equations for magnetic parameters. The number of these
quantities increases towards infinity for the full Parisi solution. 
This turned out not to be a completely unsolvable problem. At half filling we
obtained the ratio between density of states at the gap edge and gap width
as an invariant under the number of steps which break
replica permutation symmtery (RPSB)[2]. Recognizing
the gap widths to be given by the nonequilibrium susceptibility, which is
known to approach zero as $T\rightarrow0$, we concluded a pseudogap as the
exact solution. As for the Hubbard model, doping leads to serious 
complications in the fermionic spin glass. Away from half-filling the
fermionic Ising spin glass must accomodate doubly occupied sites (even at
$T=0$ and positive chemical potential).\\
As Fig.\ref{dos} indicates, the 
lowest order calculation with a fermionic excitation gap
$E_{g0}=\sqrt{2/\pi}$ keeps $\nu=1$ for $\mu<1/\sqrt{2\pi}$. 
The regions with and without a central nonmagnetic band are separated by
the line ${\cal{L}}$, shown in Fig.2. Since
$\mu=0.39$ is chosen here very close to $\frac{1}{2}E_{g0}$ and since the
line
${\cal{L}}$ includes a small area below $\frac{1}{2}E_{g0}$ at finite
temperatures,
before it turns around to meet $E_{g0}/2$ at $T=0$, a certain kind of
reentrance is observed. Decreasing the temperature one observes the growth
of a small central peak, which looses however its weight in favour of the
magnetic bands as $T\rightarrow0$.\\
Picking one representation from our numerous calculations for $\mu$ larger
than $E_{g0}/2$, where the system is no more half-filled, we selected the
one for $\mu=0.8$, a value still in the magnetic regime but not far from
the discontinuous transition into the paramagnetic phase.
For all $\mu>1/\sqrt{2\pi}$ the central peak grows until it
reaches its maximum height at zero temperature.\\
This band becomes completely separated from the magnetic bands to its left
and right, when $T$ reaches zero. The area under the central band is equal
to $\nu-1$, which is the deviation from half-filling
($\nu=\sum_{\sigma}(\hat{n}_{\uparrow}+\hat{n}_{\downarrow})$=1).
\\
We have also calculated numerically and analytically the 1-step replica
permutation symmetry broken solutions, which will be published elsewhere.
In this improved approximation a similar scenario takes place, only with
the gap energy depressed to the much smaller value of
$E_{g1}/2\approx.119$.
For chemical potentials larger than this value the system is no more half
filled.
In the range $E_{g1}<2\mu<E_{g0}$ the central peak is already present,
only
its growth is confined to lower temperatures. This will continue for more
and more RPSB-steps and, finally, the central peak will exist everywhere
except at half-filling. The exact solution for the nonmagnetic charge band
at $\infty$-RPSB maintains features of its replica symmetric Gaussian
predecessor. \\
The special line {\cal{L}} shown in Figure 2 encloses the 3-band low $T$
domain and its definition resembles an instability condition. However,
offdiagonal
elements $Q^{a\neq b}$ exist (ie finite saddle point solutions in the
infinite-range model) and stabilize the system against random field
criticality. It is the Almeida Thouless eigenvalues which decide
stability. The ubiquituous instability against RPSB does not allow exact
conclusions on other potential instabilities. This renders the task
extremely difficult and has led to delay in attempts to solve related
problems. The possibility of vector replica symmetry breaking must be
seriously considered[5]. \\
\begin{figure}
\epsfxsize=7.5cm
\epsfbox{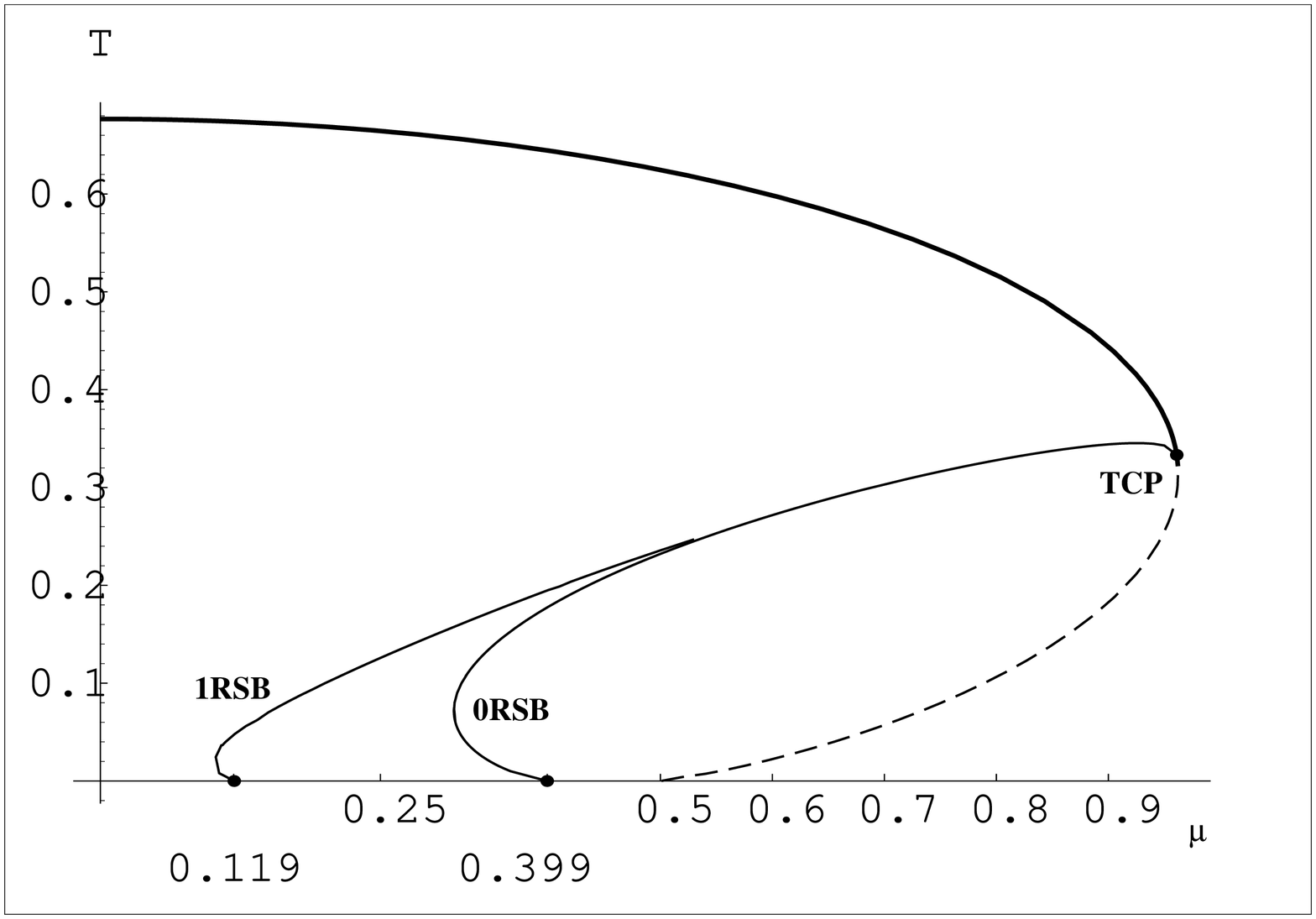}
\par\vspace{-5.3cm}\hspace{8cm}
\epsfxsize=9cm
\epsfbox{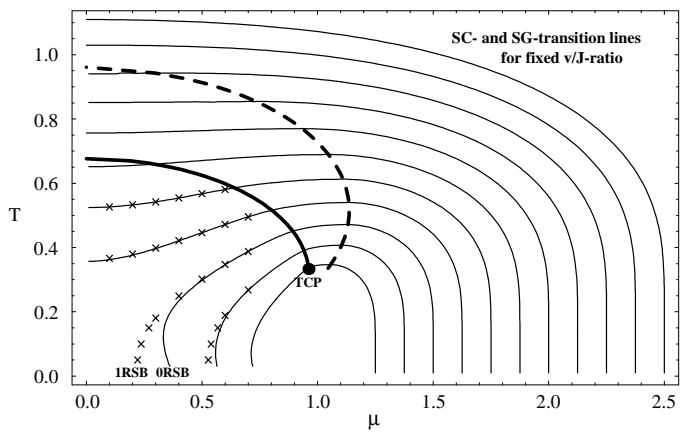}
\vspace{1cm}
\caption{\small{The spin glass phase diagram as a function of chemical
potential $\mu$ and temperature $T$ (left). A special line
${\cal{L}}$, defined by $\partial^2 F/\partial(Q^{aa})^2=0$ (F:
free energy, $Q^{aa}$: diagonal elements of the Parisi order
parameter matrix, spin autocorrelation function),
is shown in the replica symmetric approximation and in one step symmetry
breaking. It join(s) tricritical point and gap edge values at $T=0$. 
The Figure on the right shows the competition between spin glass
phase (below fat line) and local pairing (SC) phase in the ($\mu,T$)-plane
too; (thin) curves are parametrized by the ratio between attractive
interaction and frustrated magnetic coupling $J$. Crosses belong to the
1RSB-calculation}; dashed line joins tricritical SC-points.}
\label{instabline}
\end{figure}

{\bf 3. COMPETITION BETWEEN LOCAL SUPERCONDUCTING AND SPIN GLASS ORDER}\\
This type of problem was studied by several groups, who focused on models
which
couple a localized spin system with randomly frozen magnetic moments to another
species of mobile and eventually superconducting fermions. We considered
in particular the case of a single fermion species that is exposed to the
abovementioned random magnetic interaction and to an attractive
interaction. Following the experience with negative U Hubbard models for
example, arbitrarily small transport processes are known to delocalize the
preformed pairs and turn the system superconducting.\\
Many details of our analysis were published in [4]. 
The new insight gained from the selforganized 3-band structure
in the low temperature regime of the purely random magnetic system
appears relevant for local pairing and superconductivity too. The central
nonmagnetic band meets a magnetic band at the Fermi level. For any finite
order of RPSB-steps they were separated by finite gaps, but finally these 
become pseudogaps. The phase diagram shown in Fig.2 reveals the increasing
importance of RPSB for the competition between local pairing and spin
glass order as the temperature decreases. The presence of smaller and smaller order parameters in
higher RPSB order seems to allow superconductivity to
regain more pieces of the phase diagram. Very unlike the competition
between ferromagnetism and spin glass order for incomplete frustrated
interactions, the SC-SG interface does not tend towards an infinitely
steep line. Coexistence between singlet local pairing and
spin glass order parameters was not found, but the vicinity of the phases
in ($\mu,T$) led to special pairbreaking behaviour.\\
One major goal for the future is of course to find as many features as
possible of the quantum Parisi solution for the Ising type models. The
relation between quantum dynamics and replica symmetry breaking
must also be analysed in systems which are strongly quantum
dynamical in spin- and charge correlations (superconducting order
parameter adds this type of quantum dynamics too, but
amounts only to shifts of the first order transition line between
superconductor and spin glass). \\


{\bf Acknowledgments}\\
This work was supported by project Op28/5-1 and by the SFB 410 at the 
university of W\"urzburg.\\

{\bf References}

\begin{itemize}

\item[{[1]}]
S.~K. Ghatak and D. Sherrington, J. Phys. C {\bf 10},  3149  (1977).

\item[{[2]}]
R. Oppermann and B. Rosenow, Europhys. Lett. {\bf 41},  525  (1998).

\item[{[3]}]
R. Oppermann and B. Rosenow, Phys. Rev. Lett. {\bf 80},  4767  (1998).

\item[{[4]}]
H. Feldmann and R.Oppermann, Eur. Phys. J. B {\bf 10}, 429 (1999).

\item[{[5]}]
V. Dotsenko and M. M\'ezard, J. Phys. A {\bf 30},  3363  (1997).

\end{itemize}

\end{document}